\documentclass[usenatbib]{mn2e}
\usepackage{natbibmnfix,graphicx,times}
\usepackage{amssymb,amsmath}
\usepackage{epstopdf}
\newcommand{\Msun}{\mbox{ M$_\odot$}}

\newcommand{\isec}{\mbox{ s$^{-1}$}}

\newcommand{\bq}{\begin{equation}}
\newcommand{\eq}{\end{equation}}
\newcommand{\bqa}{\begin{eqnarray}}
\newcommand{\eqa}{\end{eqnarray}}

\newcommand{\lya}{Ly$\alpha$ }
\newcommand{\lyans}{Ly$\alpha$}
\newcommand{\lyb}{Ly$\beta$ }
\newcommand{\lybns}{Ly$\beta$}
\newcommand{\lyg}{Ly$\gamma$ }
\newcommand{\lygns}{Ly$\gamma$}
\newcommand{\lyn}{Ly$n$ }

\newcommand{\apj}{ApJ}
\newcommand{\apjl}{ApJ}
\newcommand{\apjs}{ApJS}

\newcommand{\aj}{AJ}
\newcommand{\mnras}{MNRAS}

\title[Lyman series cascades and the 21 cm line]{Descending from on high: Lyman series cascades and spin-kinetic temperature coupling in the 21 cm line}

\author[Pritchard and Furlanetto]{Jonathan R. Pritchard\thanks{Email: jp@tapir.caltech.edu} and Steven R. Furlanetto\\ 
Division of Physics, Mathematics, \& Astronomy, California Institute of Technology, Mail Code 130-33, Pasadena, CA 91125, USA} 

\begin{document}
\maketitle

\begin{abstract}
We examine the effect of Lyman continuum photons on the 21 cm background in the high-redshift universe. The brightness temperature of this transition is determined by the spin temperature $T_s$, which describes the relative populations of the singlet and                
triplet hyperfine states.  Once the first luminous sources appear, $T_s$ is set by          
the Wouthuysen-Field effect, in which Lyman-series photons mix the hyperfine levels.
 Here we consider coupling through $n>2$ Lyman photons.  We first show that coupling (and heating) from scattering of          
Ly$n$ photons is negligible, because they rapidly cascade to lower-energy           
photons.  These cascades can result in either a Ly$\alpha$ photon -- which          
will then affect $T_s$ according to the usual Wouthuysen-Field mechanism --            
or photons from the $2s \rightarrow 1s$ continuum, which escape without             
scattering.  
We show that a proper treatment of the cascades delays                 
the onset of strong Wouthuysen-Field coupling and affects the power                 
spectrum of brightness fluctuations when the overall coupling is still                
relatively weak (i.e., around the time of the first stars).  Cascades damp          
fluctuations on small scales because only $\sim 1/3$ of Ly$n$ photons               
cascade through Ly$\alpha$, but they do not affect the large-scale power            
because that arises from those photons that redshift directly into the              
Ly$\alpha$ transition.  We also comment on the utility of Ly$n$                  
transitions in providing ``standard rulers" with which to study the                  
high-redshift unvierse.  
\end{abstract}


\section{Introduction}
\label{sec:intro}
One potentially promising probe of the cosmic dark ages is 21 cm tomography.  It has long been known \citep{hogan_rees1979,scott_rees1990} that neutral hydrogen in the intergalactic medium (IGM) may be detectable in emission or absorption against the cosmic microwave background (CMB) at the wavelength of the redshifted 21 cm line, the spin-flip transition between the singlet and triplet hyperfine levels of the hydrogen ground state.  The brightness of this transition will thus trace the distribution of HI in the high-redshift universe \citep{field1958,field1959spint}, which gives the signal angular structure as well as structure in redshift space.  These features arise from inhomogeneities in the gas density field, the hydrogen ionization fraction and the spin temperature. \citet*{mmr1997} showed that the first stars could cause a rapid evolution in the signal through their effect on the spin temperature.  Consequently, the 21 cm signal can provide unparalleled information about the ``twilight zone" when the first luminous sources formed and the epoch of reionization and reheating commenced.

Despite the theoretical promise of this probe, it is only with improvements in computing power that building radio arrays with sufficient sensitivity, capable of correlating billions of visibility measurements, has become possible \citep{morales_hewitt2004}.  Three such arrays (LOFAR\footnote{See http://www.lofar.org/.}, MWA\footnote{See http://web.haystack.mit.edu/arrays/MWA/.}, and PAST\footnote{See \citet{pen2005past}.}) will soon be operational, opening a window onto this new low frequency band. Before a detection can be made, however, there are still major scientific and technical challenges to be met.  Ionospheric scattering and terrestrial interference are two serious issues.  Also worrying is the need to remove foregrounds, which are many orders of magnitude stronger than the signal.  Multifrequency subtraction techniques \citep*{zfh2004freq,morales_hewitt2004,santos2005}, exploiting the smoothness of the foreground spectra, have been proposed, but their effectiveness has yet to be tested.  The challenges are great, but so are the opportunities.  It is thus crucial to understand the nature of the 21 cm signal as we commence these searches.

Fluctuations in the 21 cm signal arise from both cosmological and astrophysical sources.  Most previous work has focussed on the signal due to density perturbations \citep{mmr1997,loeb_zald2004_21cm} or from inhomogeneous ionization \citep*{ciardi2003,fsh2004,fzh2004}.  An additional source of fluctuations is the spin temperature, which describes the relative occupation of the singlet and triplet hyperfine levels.  These levels may be excited by three primary mechanisms: absorption of CMB photons, atomic collisions, and absorption and re-emission of \lya photons (the Wouthuysen-Field effect;  \citealt{wouth1952,field1959spint}).  The first two processes rely upon simple physics, but the last one allows us to study the properties of luminous sources, which determine the background radiation field.

\citet[henceforth BL05]{bl2005detectgal} studied the signal generated by the first generation of collapsed objects.  These high redshift objects are highly biased, leading to large variations in their number density.   This, combined with the $1/r^2$ dependence of the flux, causes large fluctuations in the \lya background, which can be probed through their effect on the 21 cm transition.  Exploiting the anisotropy induced by peculiar velocities (\citealt{bharadwaj2004,bl2005sep}), they showed that information about the \lya radiation field could be extracted from the power spectrum of 21 cm fluctuations and separated into those  fluctuations correlated and uncorrelated with the density field.  The features of these spectra allow extraction of astrophysical parameters such as the star formation rate and bias.  However, it is not a trivial task to relate the emissivity to a distribution of \lya photons.  The background in the \lya line is composed of two parts: those photons that have redshifted directly to the \lya frequency and those produced by atomic cascades from higher Lyman series photons. To calculate this latter component, BL05 assumed that atomic cascades were $100\%$ efficient at converting photons absorbed at a Lyman resonance into a \lya photon, while in reality most cascades end in two photon decay from the $2S$ level.  

In this paper, we calculate the exact cascade conversion probabilities from basic atomic physics.  In addition, we discuss the possibility of level mixing by scattering of \lyn photons via a straightforward generalisation of the Wouthuysen-Field effect.  We then apply the cascade efficiencies to calculate the \lya flux profile of an isolated source.  The existence of discrete horizons, determined by the maximum distance a photon can travel before it redshifts into a given Lyman resonance, imprints a series of discontinuities into the profile, which can in principle be used as a standard ruler.  We apply these results to the power spectra of 21 cm fluctuations during the epoch of the first stars, showing that these corrections can not be ignored when extracting astrophysical parameters. 

The layout of this paper is as follows.  In \S \ref{sec:formalism} we introduce the formalism for describing 21 cm fluctuations and the dominant coupling mechanism, the Wouthuysen-Field effect.  In \S \ref{sec:direct} we discuss the possibility of direct pumping by Ly$n$ photons.  Next, in \S \ref{sec:cascade}, we detail the atomic physics of radiative cascades in atomic hydrogen.  The results are applied to the \lya flux profile of an isolated source in \S \ref{sec:flux} and to the 21 cm power spectrum from 
the first galaxies in \S \ref{sec:galaxies}.  We also discuss some of the limitations of this formalism.  Finally, we summarise our results in \S \ref{sec:conclusions}.  In an Appendix, we review the equations needed to calculate analytically the Einstein $A$ coefficients for the hydrogen atom.  Throughout, we assume $(\Omega_{m}, \Omega_b,\Omega_\Lambda,h,\sigma_8,n_s)=(0.3,0.046,0.7,0.7,0.9,1.0)$, consistent with the most recent measurements \citep{spergel2003wmap}.  

During the preparation of this paper, \citet{hirata2005} submitted a preprint covering similar material.  We have confirmed agreement where there is overlap.  The main results of this work were discussed at the ``Reionizing the Universe" conference in Groningen, The Netherlands (June 27-July 1, 2005; see http://www.astro.rug.nl/$\sim$cosmo05/program.html).

\section{21 cm formalism and the Wouthuysen-Field mechanism}
\label{sec:formalism}

The 21 cm line of the hydrogen atom results from hyperfine splitting of the $1S$ ground state due to the interaction of the magnetic moments of the proton and the electron.  The HI spin temperature $T_s$ is defined via the relative number density of hydrogen atoms in the $1S$ singlet and triplet levels $n_1/n_0=(g_1/g_0)\exp(-T_\star/T_s)$, where $(g_1/g_0)=3$ is the ratio of the spin degeneracy factors of the two levels, and $T_\star\equiv hc/k\lambda_{21 \rm{cm}}=0.0628\,\rm{K}$.  The optical depth of this transition is small at all relevant redshifts, so the brightness temperature of the CMB is
\begin{equation}\label{brightnessT}
T_b=\tau\left(\frac{T_s-T_{\rm{CMB}}}{1+z}\right),
\end{equation}
where the optical depth for resonant 21 cm absorption is 
\begin{equation}
\tau=\frac{3c\lambda^2hA_{10}n_{\rm{HI}}}{32\pi k_B T_s(1+z)(dv_r/dr)}.
\end{equation}
Here $n_{\rm{HI}}$ is the number density of neutral hydrogen, $A_{10}=2.85\times10^{-15}\,\rm{s}^{-1}$ is the spontaneous emission coefficient, and $dv_r/dr$ is the gradient of the physical velocity along the line of sight with $r$ the comoving distance.
When $T_s<T_{\rm{CMB}}$ there is a net absorption of CMB photons, and we observe a decrement in the brightness temperature.  

The spin temperature is determined by three coupling mechanisms.  Radiative transitions due to absorption of CMB photons (as well as stimulated emission) tend to drive $T_s\rightarrow  T_{\rm{CMB}}$.  Spin flips from atomic collisions drive $T_s\rightarrow T_k$, the gas kinetic temperature. Finally, the Wouthuysen-Field effect \citep{wouth1952,field1958}, which is the main focus of this paper, also drives $T_s\rightarrow T_k$ (see below).  The combination that appears in \eqref{brightnessT} can be written
\begin{equation}
\frac{T_s-T_{\rm{CMB}}}{T_s}=\frac{x_{\rm{tot}}}{1+x_{\rm{tot}}}\left(1-\frac{T_{\rm{CMB}}}{T_k}\right),
\end{equation}
where $x_{\rm{tot}}=x_\alpha+x_c$ is the sum of the radiative and collisional coupling parameters.  The latter is
\begin{equation}
x_c=\frac{4\kappa_{1-0}(T_k)n_HT_\star}{3A_{10}T_{\rm{CMB}}},
\end{equation}
where $\kappa_{1-0}$ is tabulated as a function of $T_k$ \citep{allison1969,zygelman2005}. The spin temperature becomes strongly coupled to the gas temperature when $x_{\rm{tot}}\gtrsim1$.

\begin{figure}
\begin{center}
\resizebox{8cm}{!}{\includegraphics{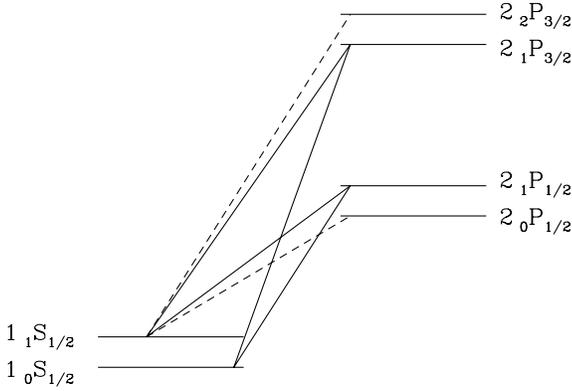}}\\%
\caption{Hyperfine structure of the $2P$ and $1S$ level of the hydrogen atom.  Levels are labelled according to the notation $n\,  {}_F L_J$, where $n$, $L$, and $J$ are the usual radial, orbital angular momentum and total angular momentum quantum numbers.  $F=I+J$ is the quantum number obtained from the nuclear spin and $J$.  Allowed transitions obey $\Delta F=0,\pm1$ (except $0\rightarrow0$).  Those relevant for the Wouthuysen-Field effect are indicated by solid curves, while dashed curves indicate the remaining allowed transitions.}
\label{fig:hyperfine}
\end{center}
\end{figure}

A schematic diagram of the Wouthuysen-Field effect is shown in Figure \ref{fig:hyperfine}; it mixes the hyperfine levels through absorption and re-emission of \lya photons. Quantum selection rules allow transitions for which the total spin angular momentum $F$ changes by $\Delta F=0,\pm1$ (except $0\rightarrow0$), making only two of the four $n=2$ levels accessible to both the $n=1$ singlet and triplet states.  Transitions to either of these states can change $T_s$.  The coupling coefficient is
\begin{equation}
x_\alpha=\frac{4P_\alpha T_\star}{27A_{10}T_{\rm{CMB}}},
\end{equation}
where $P_\alpha$ is the Ly$\alpha$ scattering rate \citep{mmr1997}.  If resonant scattering of Ly$\alpha$ photons occurs rapidly enough $T_s$ will be driven to $T_\alpha$, the colour temperature of the radiation field at the Ly$\alpha$ frequency \citep{field1958,mmr1997}.
In parallel, the repeated scattering of Ly$\alpha$ photons by the thermal distribution of atoms brings $T_\alpha\rightarrow T_k$ \citep{field1959relax,hirata2005}.  Consequently, the Wouthuysen-Field effect provides an effective coupling between the spin temperature and the gas kinetic temperature. 

We can also write the Wouthuysen-Field coupling as
\begin{equation}\label{xa_flux}
x_\alpha=\frac{16\pi^2T_\star e^2 f_\alpha}{27A_{10}T_{\rm{CMB}} m_e c}S_\alpha J_\alpha,
\end{equation}
where $f_\alpha=0.4162$ is the oscillator strength for the \lya transition, $S_\alpha$  is a correction factor of order unity \citep{chen2004,hirata2005} that accounts for the redistribution of photon energies due to repeated scattering off the thermal distribution of atoms, and $J_\alpha$ is the angle-averaged specific intensity of \lya photons by photon number.  For reference, a \lya flux of $J_\alpha=1.165\times10^{-10}\left[(1+z)/20\right]\rm{cm}^{-2}\isec\,\rm{Hz}^{-1}\,\rm{sr}^{-1}$ yields $x_\alpha=1$ (corresponding to $P_\alpha=7.85\times10^{-13}[1+z]\,\isec$). 

Fluctuations in the brightness temperature arise from fluctuations in the density, the Wouthuysen-Field coupling, the neutral fraction $x_{\rm{HI}}$ and the radial velocity component.  To linear order
\begin{equation}\label{deltatb}
\delta_{T_b}=\beta\delta+\frac{x_\alpha}{\tilde{x}_{\rm{tot}}}\delta_{x_\alpha}+\delta_{x_{\rm{HI}}}-\delta_{d_r v_r},
\end{equation}
where $\delta_a$ is the fractional perturbation in $a$, $\delta$ is the fractional density perturbation, and $\tilde{x}_{\rm{tot}}=x_{\rm{tot}}(1+x_{\rm{tot}})$.  $\beta$ is a parameter describing the thermal history of the gas, which we assume to have cooled adiabatically, so that $\beta\approx0.2$.  \citet{naoz2005} showed that $\beta$ slightly increases and exhibits mild scale dependence, as gas temperature fluctuations do not exactly track the density fluctuations.  We note that, on scales $0.01\,{\rm Mpc^{-1}}<k<10^3\,{\rm Mpc^{-1}}$, $\beta$ is approximately constant at $z=20$ and choose to ignore this subtlety for ease of comparison with BL05.  The first three components of equation \eqref{deltatb} are isotropic, but the velocity fluctuation introduces an anisotropy of the form $\delta_{d_r v_r}(k)=-\mu^2\delta$ in Fourier space \citep{bharadwaj2004}, where $\mu$ is the cosine of the angle between the wavenumber $\boldsymbol{k}$ of the Fourier mode and the line of sight.  This allows us to separate the brightness temperature power spectrum $P_{T_b}$ into powers of $\mu^2$ \citep{bl2005sep}
\begin{equation}\label{powertb}
P_{T_b}(\boldsymbol{k})=\mu^4P_{\mu^4}(\boldsymbol{k})+\mu^2P_{\mu^2}(\boldsymbol{k})+P_{\mu^0}(\boldsymbol{k}).
\end{equation}
The anisotropy is sourced only by density fluctuations, so that $P_{\mu^4}$ depends only on the matter power spectrum. $P_{\mu^2}$ contains cross-correlations between matter fluctuations and both $\delta_{x_\alpha}$ and $\delta_{x_{\rm{HI}}}$, making it an ideal probe of fluctuations in the radiation background.  In particular, at sufficiently high redshifts such that $x_{\rm{HI}}\ll 1$, it probes variations in the \lya background. Linear combinations of these three terms can be used to extract detailed information about other types of fluctuations \citep{bl2005sep}.

\section{Direct pumping by Lyman series photons}
\label{sec:direct}

Of course, the radiation background contains photons that redshift into all the Lyman transitions, not just \lyans.  The main purpose of this paper is to examine how these affect $T_s$.  The existing literature assumes that all \lyn photons are immediately converted into \lya photons by atomic cascades (e.g. BL05).  In reality, there are two different contributions to consider: one due to scattering of the \lyn photon itself and the other due to its cascade products.  In this section, we discuss the direct contribution of \lyn scattering to the coupling of $T_s$ and $T_k$, which occurs in a manner exactly analogous to the Wouthuysen-Field effect.  For this effect to be significant two requirements must be fulfilled.  First, the scattering rate of Ly$n$ photons must be sufficient to couple $T_s$ and $T_n$, the Ly$n$ colour temperature.  Second, it must be sufficient to drive $T_n\rightarrow T_k$.  We will argue that neither condition is satisfied in practice.

The IGM is optically thick $\tau \gg 1$ to all Lyman series transitions with $n\lesssim 100$.  Consequently, a Ly$\alpha$ photon emitted by a star will scatter many times ($\sim\tau\sim10^6$ ; \citealt{gp1965}) before it finally escapes by redshifting across the line width; each of these scatterings contributes to the Wouthuysen-Field coupling.  A Ly$n$ photon can escape by redshifting across the line width, but a transition to a level other than $n=1$ will also remove it.  The probability for a decay from an initial state $i$ to a final state $f$ is given in terms of the Einstein $A_{if}$ coefficients by
\begin{equation}\label{eqn:atomprob}
P_{if}=\frac{A_{if}}{\sum_f A_{if}}.
\end{equation}
Appendix \ref{app:einsteinA} summarises the expressions needed to compute the Einstein $A_{if}$ coefficients.
For the Lyman series transitions $P_{nP\rightarrow1S}\approx 0.8$ (see Table 1) so that a Ly$n$ photon will scatter of order $N_{\rm{scat}}\approx1/(1-P_{nP\rightarrow1S})\sim5$ times before undergoing a cascade.  

Because a cascade occurs long before escape via redshifting, the coupling from direct pumping is negligible.  Recall that the scattering rate $P_X$ for the photon type $X=$\lyans, \lybns, etc. may be expressed as \citep{field1959spint}
\begin{equation}\label{nscat}
n_{\rm{HI}} P_X=N_{\rm{scat}} \dot{n}_X
\end{equation}
in terms of the production rate of photons per unit volume $\dot{n}_X$.  It is then clear that, for similar production rates (i.e., for sources with a reasonably flat spectrum),  $P_n/P_\alpha\sim N_{\rm{scat},n}/N_{\rm{scat},\alpha} \sim 5\times10^{-6}$.  This simple argument shows that the contribution from direct pumping by Ly$n$ photons will be negligible compared to that of the Ly$\alpha$ photons, because $x_\alpha\propto P_\alpha$.

The second question, whether $T_n\rightarrow T_k$, is still relevant for heating of the gas by repeated scatterings.  Given the reduced number of scattering events, it seems unlikely to be the case, but a full calculation using a Monte Carlo method or following \citet{chen2004} is required to rigourously answer this question.  Lack of equilibrium would make \lyn scattering a more efficient source of heat, on a per scattering basis.  \citet{chen2004} have shown that Ly$\alpha$ heating is much smaller than previous calculations indicated \citep{mmr1997}, because $T_\alpha\approx T_k$, which reduces the heat transferred per collision. This is unlikely to be the case for the Ly$n$.

Following \citet{mmr1997}, we can estimate the \emph{maximum} heating from a single Ly$n$ scattering by assuming that all of the atomic recoil energy for a stationary atom is deposited in the gas.  Momentum conservation then demands
\begin{equation}
\dot{E}_n=-\left\langle\frac{\Delta E}{E}\right\rangle h\nu_n P_n,
\end{equation}
where $\langle \Delta E/E\rangle\sim 10^{-8}$ is the fraction of energy lost by a Ly$n$ photon after scattering from a stationary hydrogen atom, and $h \nu_n$ is the energy of the photon.    Assuming the production rate of Ly$n$ photons is comparable to that of the \lya photons and taking $x_\alpha=1$, we then obtain $\dot{E}_n\sim0.002[(1+z)/10]\,\rm{K\,Gyr^{-1}}$.  This is much smaller than the \lya heating rate, even including the $T_\alpha\approx T_k$ correction, so we do not expect \lyn scattering to be a significant heat source.  Furthermore, if $T_n\approx T_k$ the rate would be much smaller than this estimate, as in \citet{chen2004}.

\section{Lyman series cascades}
\label{sec:cascade}

An excited state of hydrogen may reach the ground state in three ways.  Firstly, it may decay directly to the ground state from an $nP$ state ($n>2$), generating a \lyn photon.  Secondly, it may cascade to the metastable $2S$ level.  Decay from the $2S$ level proceeds via a forbidden two photon process.  Finally, it may cascade to the $2P$ level, from which it will produce a \lya photon.  We are primarily interested in the fraction of decays that generate \lya photons, which will increase the \lya flux pumping the hyperfine levels.  

\begin{figure} 
\begin{center}
\resizebox{8cm}{!}{\includegraphics{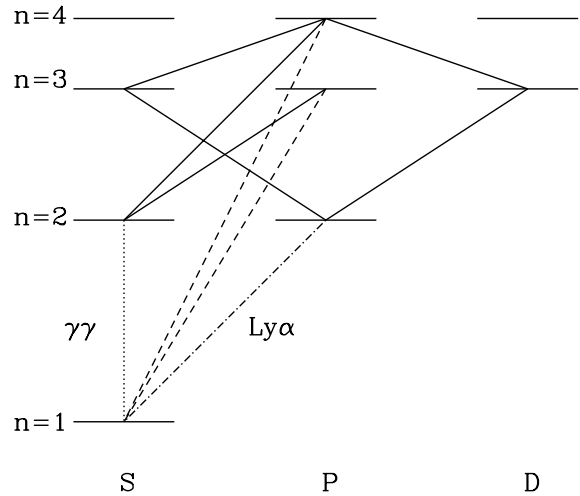}}\\%
\caption{Energy level diagram for the hydrogen atom illustrating \lyb and \lyg cascades.  Marked decays are distinguished as cascades (solid curves), \lyn (dashed curves), \lya (dot-dashed curve), and the two photon decay (dotted curve).  Note that the selection rules ($\Delta L=\pm1$)  decouple the $3P$ and $2P$ levels, preventing \lyb from being converted into \lyans.}
\label{fig:cascade}
\end{center}
\end{figure}

The fraction of cascades that generate \lya photons can be determined straightforwardly from the selection rules and the decay rates.  As an example, consider the \lyb system.  Absorption of a \lyb photon excites the atom into the $3P$ level.  As illustrated in Figure \ref{fig:cascade}, the $3P$ level can decay directly to the ground state, regenerating the \lyb photon, or to the $2S$ level, where it will decay by two photon emission.  The selection rules forbid \lyb photons from being converted into \lya photons.  In contrast, the $4P$ level, excited by absorption of \lygns, can cascade via the $3S$ or $3D$ levels to the $2P$ level and then generate \lyans.  

To calculate the probability $f_{\rm{recycle}}$ that a \lyn photon will generate a \lya photon, we apply an iterative algorithm.  The expression
\begin{equation}\label{f_recycle_iterate}
f_{\rm{recycle}\emph{,i}}=\sum_f P_{if}f_{\rm{recycle}\emph{,f}}
\end{equation}
relates the conversion probability for the initial level $i$ to the conversion probabilities of all possible lower levels $f$.  The decay probabilities are calculated using equation \eqref{eqn:atomprob}. We then iterate from low to high $n$, calculating each $f_{\rm{recycle}}$ in turn.

In our particular case of an optically thick medium, we can ignore direct transitions to the ground state.  These generate a \lyn photon, which will rapidly be reabsorbed and regenerate the $nP$ state.  Therefore, such decays will not affect the net population of photons or of excited states.  We incorporate this into the calculation by setting $A_{nP\rightarrow1S}=0$ \citep{furlane2005structure}.

\begin{figure} 
\begin{center}
\resizebox{8cm}{!}{\includegraphics{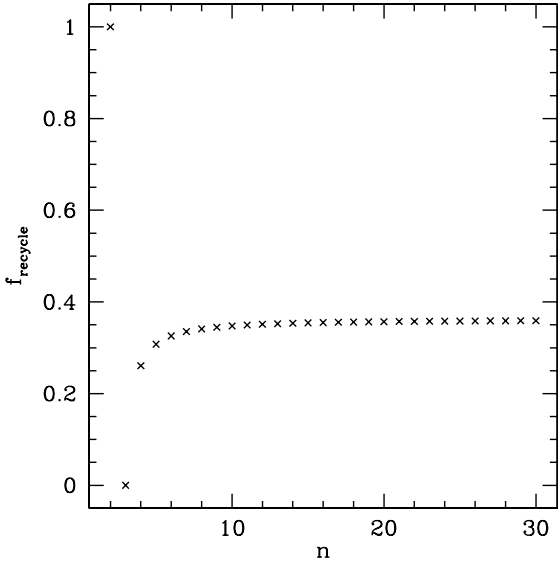}}\\%
\caption{Recycling fractions for \lyn photons.  Note that the values level off at $f_{\rm{recycle}}\approx0.359$ and that none of the photons incident on the Ly$\beta$ resonance are converted into Ly$\alpha$ photons.}
\label{fig:recycle}
\end{center}
\end{figure}

Results for the lowest Lyman series transitions are summarised in Table 1 and plotted in Figure \ref{fig:recycle}.\footnote{Code for calculating the conversion factors is available at http://www.tapir.caltech.edu/$\sim$jp/cascade/.}  These results are in agreement with those of \citet{hirata2005}.  At large $n$, the conversion fractions asymptote to $f_{\rm{recycle}}\approx0.36$ because nearly all cascades pass through lower levels.  We emphasise again that the quantum selection rules forbid a Ly$\beta$ photon from producing a Ly$\alpha$ photon.  
\begin{table}
\begin{center}
\begin{tabular}{|c|c|c||c|c|c|}
\hline
$n$ & $f_{\rm{recycle}}$ &$P_{nP\rightarrow1S}$& $n$ & $f_{\rm{recycle}}$ &$P_{nP\rightarrow1S}$\\
\hline
 &  & & 16 & 0.3550 & 0.7761\\
2 & 1 & 1 & 17 & 0.3556 & 0.7754\\
3 & 0 & 0.8817 & 18 & 0.3561 & 0.7748\\
4 & 0.2609 & 0.8390 & 19 & 0.3565 & 0.7743\\
5 & 0.3078 & 0.8178 & 20 & 0.3569 & 0.7738\\
6 & 0.3259 & 0.8053 & 21 & 0.3572 & 0.7734\\
7 & 0.3353 & 0.7972 & 22 & 0.3575 & 0.7731\\
8 & 0.3410 & 0.7917 & 23 & 0.3578 & 0.7728\\
9 & 0.3448 & 0.7877 & 24 & 0.3580 & 0.7725\\
10 & 0.3476 & 0.7847 & 25 & 0.3582 & 0.7722\\
11 & 0.3496 & 0.7824 & 26 & 0.3584 & 0.7720\\
12 & 0.3512 & 0.7806 & 27 & 0.3586 & 0.7718\\
13 & 0.3524 & 0.7791 & 28 & 0.3587 & 0.7716\\
14 & 0.3535 & 0.7780 & 29 & 0.3589 & 0.7715\\
15 & 0.3543 & 0.7770 & 30 & 0.3590 & 0.7713\\
\hline
\end{tabular}
\end{center}
\label{tab:recycle}
\caption{Recycling fractions $f_{\rm{recycle}}$ and decay probabilities to the ground state, $P_{nP\rightarrow1S}$.}
\end{table}

Finally we briefly comment on the two photon decay from the $2S$ level (for a more detailed discussion, see \citealt{hirata2005}).  Selection rules forbid electric dipole transitions from the $2S$ level to the ground state, but the second order two photon decay process can occur with $A_{\gamma\gamma}=8.2\isec\ll A_{2P\rightarrow1S}$.  At $z\lesssim 400$ the CMB flux density is sufficiently small that radiative excitations from the $2S$ level are negligible.  Additionally, at the relevant densities collisional excitation to the $2P$ level is slow compared to the two photon process \citep{breit1940}.  Consequently, the $2S$ level will preferentially decay via this two photon process.  These transitions may themselves affect $T_s$, because both the $2S$ and $1S$ levels have hyperfine structure and any imbalance in the decay constants here would affect the $1S$ populations.  However, even without detailed calculations, we can see that the resultant coupling must be small.  Cascades that do not generate \lya must reach the $2S$ level, so the fraction of \lyn photons that undergo two photon decay is $f_{\gamma\gamma}(n)=1-f_{\rm{recycle}}(n)\approx0.64$ (See Table 1).  Each such decay has $N_{\rm{scat},\gamma\gamma}=1$ because the resulting photons are not reabsorbed.  This is much smaller than $N_{\rm{scat},\alpha}\approx10^6$.  Consequently, only if the coupling per scattering were many orders of magnitude larger for two photon decay than for \lya scattering could this effect be significant.

\section{The \lya coupling around a source}
\label{sec:flux}
We can see the effects of these recycling fractions on the Ly$\alpha$ coupling by considering the life of a photon emitted from a given source.  The photon initially propagates freely, redshifting until it enters a \lyn resonance.  Because the IGM is so optically thick, the photon will then scatter several times until a cascade converts it into a Ly$\alpha$ photon or two $2S\rightarrow1S$ photons.  In the latter case, the photons escape to infinity; in the former case, it scatters $\sim\tau$ times before redshifting out of the \lya resonance.  This establishes a series of closely-spaced horizons, because a photon entering the Ly$n$ resonance at $z$ must have been emitted below a redshift 
\begin{equation}
1+z_{\rm{max}}(n)=(1+z)\frac{[1-(n+1)^{-2}]}{(1-n^{-2})}.
\end{equation}  
The number of \lyn transitions contributing \lya photons is thus a function of the distance from the source.  These horizons imprint well-defined atomic physics onto the coupling strength by introducing a series of discontinuities into the Ly$\alpha$ flux profile of a source.

Thus the \lya flux, $J_\alpha$, arises from a sum over the Ly$n$ levels, with the maximum $n$ determined by the distance.  The sum is ultimately truncated at $n_{\rm{max}}\approx23$ to exclude levels for which the horizon lies within the HII region of a typical (isolated) galaxy, as only neutral hydrogen contributes to 21 cm absorption (BL05).  The average \lya background is thus
\begin{multline}\label{jflux}
J_\alpha(z)=\sum_{n=2}^{n_{\rm{max}}}J^{(n)}_\alpha(z)\\=\sum_{n=2}^{n_{\rm{max}}}\int_z^{z_{\rm{max}}(n)} dz'\,f_{\rm{recycle}}(n)\frac{(1+z)^2}{4\pi}\frac{c}{H(z')}\epsilon(\nu'_n,z'),
\end{multline}
where $\nu'_n$ is the emission frequency at $z'$ corresponding to absorption by the level $n$ at $z$
\begin{equation}
\nu_n'=\nu_n\frac{(1+z')}{(1+z)},
\end{equation}
and $\epsilon(\nu,z)$ is the comoving photon emissivity (defined as the number of photons emitted per unit comoving volume, per proper time and frequency, at frequency $\nu$ and redshift $z$).  To calculate $\epsilon(\nu,z)$, we follow the model of BL05 
\begin{equation}
\epsilon(\nu,z)=\bar{n}^0_b f_{*}\frac{d}{dt}F_{\rm{gal}}(z)\epsilon_b(\nu),
\end{equation}
 where $\bar{n}^0_b$ is the cosmic mean baryon number density today, $f_{*}$ is the efficiency with which gas is converted into stars in galactic halos (and with which Lyman-continuum photons escape their hosts), and $F_{\rm{gal}}(z)$ is the fraction of gas inside galaxies at $z$.
We model the spectral distribution function of the sources $\epsilon_b(\nu)$ as a separate power law $\epsilon_b(\nu)\propto\nu^{\alpha_s-1}$ between \lya and \lyb and between \lyb and the Lyman limit.  In our calculations, we will assume Population III stars with spectral index $\alpha_s=1.29$ between \lya and \lybns, normalised to produce 4800 photons per baryon between \lya and the Lyman limit, of which 2670 photons are emitted between \lya and \lybns.  In contrast, for Population II stars the numbers are 0.14, 9690, and 6520 respectively (BL05).  In calculating $F_{\rm{gal}}$, we use the \citet{sheth1999mfn} mass function $dn/dm$, which matches simulations better than the \cite{ps1974mfn} mass function, at least at low redshifts.  We assume that atomic hydrogen cooling to a viral temperature $T_{\rm{vir}}\approx 10^4\,\rm{K}$ sets the minimum halo mass.   We will normalise $f_*$ so that $x_\alpha=1$ at $z=20$; this yields $f_*=0.16\%$ when we include the correct $f_{\rm{recycle}}$.

To see what fraction of photons from a given source are converted into Ly$\alpha$, we integrate  $\epsilon_b(\nu)$ with the proper weighting by $f_{\rm{recycle}}$.  We find that $\bar{f}_{\rm{recycle}}=0.63$, $0.72$, and $0.69$ for $\alpha_s(\rm{Ly}\alpha-\rm{Ly}\beta)=1.29$, $0.14$, and $-1.0$ respectively (roughly corresponding to Pop. III stars, low metallicity Pop II stars, and quasars; \citealt{zheng1997quasar}).  The total flux is significantly less than if $f_{\rm{recycle}}=1$, as has been generally assumed before.  Thus \lya coupling will take place later if the proper atomic physics are included (typically $\Delta z\gtrsim 1$ for fixed source parameters; see also \citealt{hirata2005}).  Of course, this is only the average value, and around a given source there will be a distance dependence.  A gas element that can only be reached by photons redshifted from below the Ly$\beta$ resonance will see an effective $f_{\rm{recycle}}=1$.  In contrast, a gas element very close to the source will have $f_{\rm{recycle}}\approx0.36$.  This will be reflected in the brightness temperature power spectrum (see \S \ref{sec:galaxies}).

\begin{figure} 
\begin{center}
\resizebox{8cm}{!}{\includegraphics{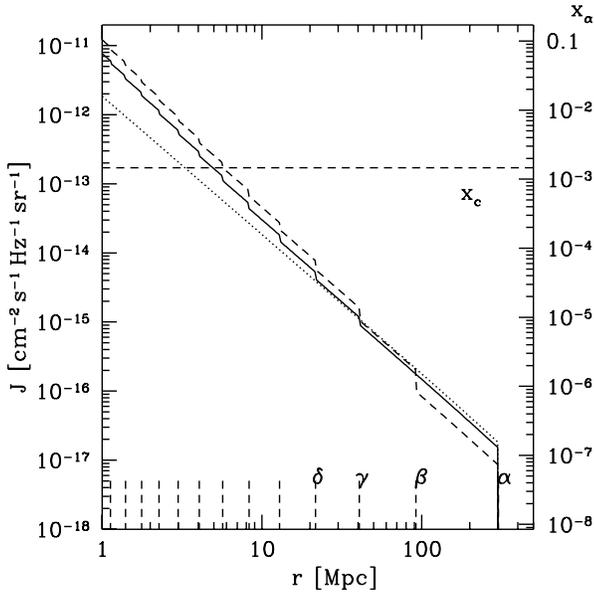}}\\%
\caption{Flux profile of a galaxy of mass $M=3\times10^{10}\Msun$ at $z=20$ as a function of comoving distance $r$.  For comparison, we have plotted the flux profile assuming $f_{\rm{recycle}}=1$ (dashed curve), proper atomic physics (solid curve) and including only photons with $\nu_\alpha<\nu<\nu_\beta$ (dotted curve).  Vertical lines along the lower axis indicate the horizons for the \lyn resonances.  The horizontal dashed line shows the value of $x_c$ at $z=20$, illustrating the regime where collisional coupling dominates.}
\label{fig:fluxprofile}
\end{center}
\end{figure}

The \lya flux profile of a galaxy with $M_{\rm{gal}}=3\times10^{10}\Msun$ and our fiducial parameters at $z=20$ is plotted in Figure \ref{fig:fluxprofile}.  In our approximation, $J_\alpha\propto M_{\rm{gal}}$.  Thus obtaining $x_\alpha\ge 1$ at $r=10\,\rm{Mpc}$ requires a galaxy mass of $M_{\rm{gal}}=4.2\times10^{12}\Msun$, corresponding to a $14\sigma$ fluctuation in the density field.  Obviously, individual sources do not induce strong Lyman coupling on large scales.  The conversion of photons from \lyn to \lya steepens the flux profile beyond the simple $1/r^2$ form.  Notice that we have normalised $f_*$ for each curve separately, so that $x_\alpha=1$ at $z=20$.  Because setting $f_{\rm{recycle}}=1$ weights large $n$ transitions more heavily (and hence small scales), that curve lies below the others at large $r$.  The discontinuities occur at the \lyn horizons.  In theory, their positions yield standard rulers determined by simple atomic physics.  In practice, the weakness of the discontinuities, and the overlapping contributions of other nearby sources, makes it unlikely that these discontinuities will be observable for an isolated source (see also \S \ref{sec:galaxies}).  Finally, we note that sharp discontinuities only occur if a photon undergoes a cascade immediately after entering a \lyn resonance and if the resulting \lya photons redshift out of the \lya resonance immediately.  The former is certainly true, but the latter will affect the shape significantly \citep{loeb_rybicki1999}.

Additionally, photon horizons affect the radiation heating of the gas around a source. The cascades, which result from scattering of \lyn photons, deposit some of their radiative energy into the kinetic energy of the gas, and their total heating rate differs from ``continuum" \lya photons because they are injected as line photons \citep{chen2004}.  Thus the \lya heating profile will differ from $1/r^2$.  However, \lya heating is typically much smaller than other sources, so this is unlikely to be important \citep{chen2004}.

\section{Brightness fluctuations from the first galaxies}
\label{sec:galaxies}

In the previous section, we saw that the proper $f_{\rm{recycle}}$ affects the spatial distribution of $x_\alpha$ around each source.  The most important manifestation of this will occur when the Wouthuysen-Field effect is just becoming important, around the time of the first galaxies (BL05).  Those authors showed that \lyn transitions enhance the small-scale fluctuations in $T_b$, but they assumed that $f_{\rm{recycle}}=1$.  We will show how the scale-dependent $f_{\rm{recycle}}$ modify this signal. It is possible to exploit the separation of powers to probe separately fluctuations that correlate with the density field and those, like Poisson fluctuations, that do not.  We consider each in turn and compare to the results of BL05.  We set $\delta_{x_{\rm{HI}}}=0$ throughout.  We also assume that the IGM cools adiabatically, with no heat input from X-rays.  Note that for ease of comparison with BL05, we do not incorporate the low-temperature corrections of \citet{hirata2005} (and in any case they are small in our example).

\subsection{Density fluctuations}
\label{ssec:density}

Density perturbations source $x_\alpha$ fluctuations via three effects (BL05).  First, the number of galaxies traces, but is biased with respect to, the underlying density field.  As a result an overdense region will contain a factor $[1+b(z)\delta]$ more sources, where $b(z)$ is the (mass-averaged) bias, and will have a larger $x_\alpha$.  Next, photon trajectories near an overdense region are modified by gravitational lensing, increasing the effective area by a factor $(1+2 \delta/3)$.  Finally, peculiar velocities associated with gas flowing into overdense regions establish an anisotropic redshift distortion, which modifies the width of the region corresponding to a given observed frequency.  These three effects may be represented using a linear transfer function $W(k)$ relating fluctuations in the coupling $\delta_{x_\alpha}$ to the overdensity $\delta$
\begin{equation}
\delta_{x_\alpha}\equiv W(k)\delta.
\end{equation}
We compute $W(k)$ for a gas element by adding the coupling due to \lya flux from each of the \lyn resonances (BL05)
\begin{multline}\label{wk}
W(k)=\frac{1}{x_\alpha}\sum_{n=2}^{n_{\rm{max}}}\int^{z_{\rm{max}}(n)}_z dz' \frac{dx^{(n)}_\alpha}{dz'}\frac{D(z')}{D(z)}\\ \times \left\{[1+b(z')]j_0(kr)-\frac{2}{3}j_2(kr)\right\},
\end{multline}
where $D(z)$ is the linear growth function and the $j_l(x)$ are spherical Bessel functions of order $l$.  The first term in brackets accounts for galaxy bias while the second describes velocity  effects. The ratio $D(z')/D(z)$ accounts for the growth of perturbations between $z'$ and $z$. The factor $dx_\alpha/dz$ converts from \lya flux to the coupling. Each resonance contributes a differential coupling (see eq. \ref{xa_flux})
\begin{equation}\label{xasum}
\frac{dx^{(n)}_\alpha}{dz'}\propto\frac{dJ_\alpha^{(n)}}{dz'},
\end{equation}
with the differential comoving flux in \lya from equation \eqref{jflux}.

\begin{figure} 
\begin{center}
\resizebox{8cm}{!}{\includegraphics{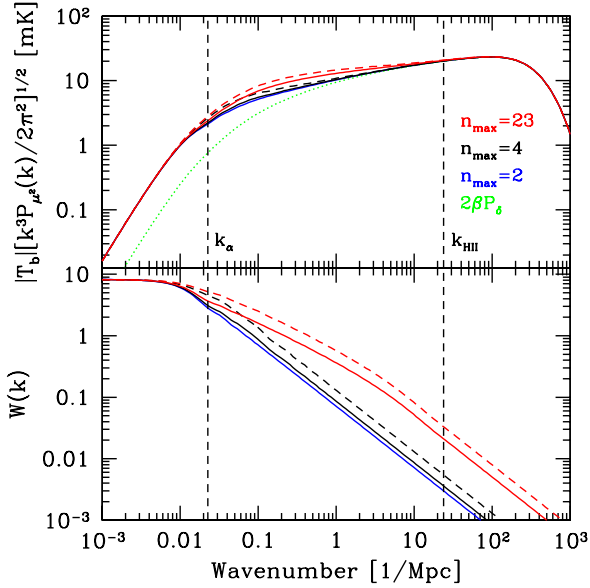}}\\%
\caption{\emph{Top panel}: $P_{\mu^2}$ power spectrum, which best illustrates the $T_b$ fluctuations arising from density-sourced $x_\alpha$.  From bottom to top, the cases include only photons from Ly$\alpha$ to Ly$\beta$, from Ly$\alpha$ to Ly$\delta$, and for $n\le23$.  Dashed lines indicate $f_{\rm{recycle}}=1$ while solid lines use the proper conversion factors. The dotted line is $2\beta P_\delta$.  Vertical lines indicate the scales corresponding to the Ly$\alpha$ horizon $r_\alpha$ and the HII region size $r_{\rm{HII}}$.  \emph{Bottom panel}: Transfer function $W(k)$.}
\label{fig:densityplot}
\end{center}
\end{figure}

Because this correlates with the density field, it is easiest to observe via
\begin{equation}\label{pmu2}
P_{\mu^2}(k)=2P_\delta(k)\left[\beta+\frac{x_\alpha}{\tilde{x}_{\rm{tot}}}W(k)\right].
\end{equation}
The first term probes fluctuations in $T_k$ and $\kappa_{1-0}$ (all encoded in $\beta$).  We show $P_{\mu^2}$ in Figure \ref{fig:densityplot},\footnote{Our results for $\sqrt{P_{\mu^2}}$ are a factor of $\sqrt{2\pi^2}\approx4.4$ greater than those of BL05, who made an error in their $P_\delta(k)$ normalisation (R. Barkana, private communication).  Note that all of their density-induced fluctuation amplitudes should increase by a similar amount.} contrasting cases that include only photons with $\nu_\alpha<\nu<\nu_\beta$, $\nu_\alpha<\nu<\nu_\delta$, and the entire Lyman continuum.  For the latter two models, we show results with $f_{\rm{recycle}}=1$ and with the proper atomic physics.  Note that each is separately normalised to $x_\alpha=1$.  The dotted curve isolates $2P_\delta\beta$, which clearly dominates on small scales.  Note that we have applied two cutoffs to the power spectrum in this regime (BL05).  The first is due to baryonic pressure, which prevents collapse on small scales.  The second is the thermal width of the 21 cm line.  \citet{naoz2005} showed that this thermal cutoff displays a characteristic angular dependence, which, theoretically, allows fluctuations and the cutoff to be separated.  To allow easy comparison with BL05, we do not include this refinement.  Additionally, we expect power from the HII regions surrounding the sources to become important on scales smaller than the size of a typical HII region $r_{\rm{HII}}$.  We have marked this scale for an isolated galaxy in Figure \ref{fig:densityplot}, but note that \citet{fzh2004} predict that the HII regions could be a factor of a few larger at these early times. On sufficently large scales $kr\approx0$, the second term in equation \eqref{pmu2} dominates, and $W(k)$ is fixed by the source bias.

\begin{figure} 
\begin{center}
\resizebox{8cm}{!}{\includegraphics{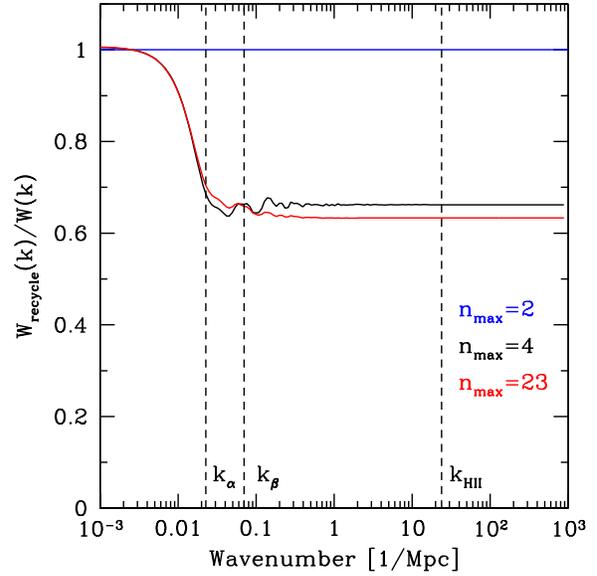}}\\%
\caption{Ratio of $W(k)$ calculated using proper atomic physics to $W(k)$with $f_{\rm{recycle}}=1$.  From left to right, vertical lines indicate the scales associated with the \lya and \lyb resonances and with $r_{\rm{HII}}$.  $W(k)$ displays small amplitude ripples, which arise from integrating an oscillating kernel (the spherical Bessel functions in equation \ref{wk}) over finite extent (the \lyn horizons).  Changing $f_{\rm{recycle}}$ modifies the phase of these ripples leading to the wiggles seen in $W_{\rm{recycle}}(k)/W(k)$ on intermediate scales.}
\label{fig:wkplot_rel}
\end{center}
\end{figure}
Figure \ref{fig:densityplot} clearly shows that the \lyn resonances are important on intermediate scales.  On large scales only the average flux matters, but as we move to smaller scales the higher-$n$ levels become important.  Figure \ref{fig:wkplot_rel} shows that the fractional reduction in $W(k)$ on small scales is $\sim0.63$.  Although this is near $\bar{f}_{\rm{recycle}}$ for Pop. III stars, that is not the origin of this scaling.  In equation \eqref{wk}, we can write
\begin{equation}
x_\alpha=\sum_n a_n f_{\rm{recycle}}(n),
\end{equation}
and
\begin{equation}
W(k)=\frac{1}{x_\alpha}\sum_n b_n f_{\rm{recycle}}(n),
\end{equation}
where the $b_n$ and $a_n$ are defined by reference to equation \eqref{wk} and the integral of equation \eqref{xasum} respectively.  The former care only about the local flux, but the latter are averaged over the entire Lyman continuum.
If $f_{\rm{recycle}}=\rm{constant}$, they cancel out of $W(k)$ and are relevant only as an overall normalisation of $x_\alpha$.  In actuality, $f_{\rm{recycle}}$ is a function of $n$, reducing the power.  We stress that this is because the $f_{\rm{recycle}}(n)$ are essentially frequency dependent and so distort the flux profile about any isolated galaxy (see Fig. \ref{fig:fluxprofile}).  

It might be hoped that the discontinuities in Figure \ref{fig:fluxprofile} would leave a clear feature on the power spectrum, especially one associated with the loss of all photons entering the Ly$\beta$ resonance.  Such a feature, whose angular scale would be determined by simple atomic physics, could set a standard ruler that could be used to test variations in fundamental constants or to measure cosmological parameters.  Sadly, as can be seen from Figure \ref{fig:densityplot}, there is no truly distinct feature. Still, the power does decline around $k_\alpha$ and measuring its shape can constrain the angular diameter distance: we find that the amplitude of $P_{\mu^2}$ changes by a few percent if the angular diameter distance changes by the same amount.  However, such constraints would also require the astrophysical parameters to be known precisely, which will be difficult.

\subsection{Poisson fluctuations}
\label{ssec:poisson}

We turn now to brightness fluctuations uncorrelated with the underlying density perturbations, which can be extracted from the power spectrum because of the redshift space distortions (BL05).  Specifically, if the number density of galaxies is small, then Poisson fluctuations can be significant.  To calculate the correlation function from Poisson fluctuations, we again follow BL05 and consider the \lya flux from sources within a volume element $dV$ at two points A and B separated by a comoving distance $l$.  The correlation function takes the form
\begin{equation}\label{eqn:poissoncorrfn}
\xi_P(l)=\frac{2}{x_\alpha^2}\int_V dV\int_M \frac{dn(z_A')}{dM} dM M^2 \frac{P(z_A')}{r_A^2}\frac{P(z_B')}{r_B^2}\frac{F_{\rm{gal}}(z_B')}{F_{\rm{gal}}(z_A')},
\end{equation}
where $z_B'=z'(r_B)$ is the redshift of a halo at a comoving distance $r_B$ from a gas element at redshift $z$.  In this expression, we integrate over a half volume such that $r_A<r_B$, with the factor of 2 accounting for the contribution of sources that are nearer to B.  The factors $P(z')$ serve to normalise the flux from $dV$ such that
\begin{equation}
dx_\alpha\equiv P(z')\frac{1}{r^2}\int_M M\frac{dn(z')}{dM}dMdV,
\end{equation}
which makes explicit the expected $1/r^2$ dependence of the flux.  Because of the finite speed of light, points A and B see the sources within $dV$ at different stages in their evolution.  Following BL05, we account for this with the last factor, which scales the source flux by the fraction of mass that has collapsed at the observed redshift.  This ignores a possible dependence of the formation rate on halo mass, but in practice  high redshift galaxies are highly biased and occur within a small mass range just above the minimum cooling mass, so this dependence will be weak.  Equation \eqref{eqn:poissoncorrfn} is easy to understand.  For Poisson statistics, the variance of flux from a set of identical galaxies would be $\propto m_{\rm{gal}}^2n_{\rm{gal}}V$; this must then simply be weighted by the flux reaching each of the two points.  Note also that, contrary to the claims of BL05, $\xi_P\propto 1/f_d$, where $f_d$ is the duty cycle of each galaxy, because the fluctuations are weighted by two powers of luminosity $[MP(z)]$ but only one factor of the density.
\begin{figure} 
\begin{center}
\resizebox{8cm}{!}{\includegraphics{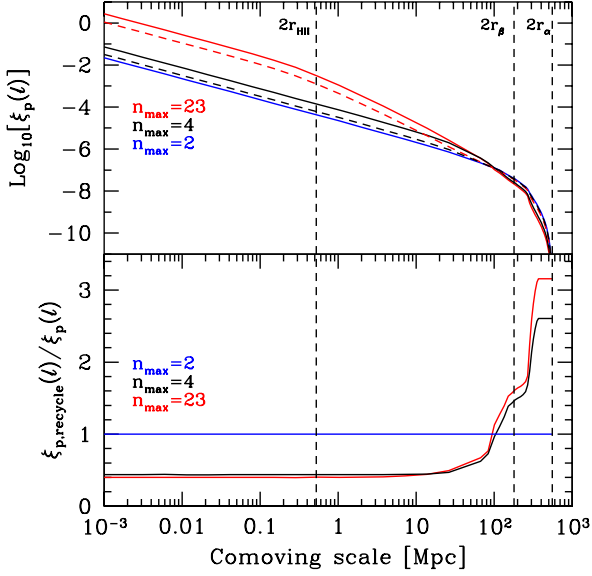}}\\%
\caption{\emph{Top panel}: Correlation function for the Poisson fluctuations.  Line conventions are the same as in Figure \ref{fig:densityplot}.  Vertical lines show $2r_\alpha$, $2r_\beta$, and $2r_{\rm{HII}}$. \emph{Bottom panel}: Ratio of the correlation functions assuming $f_{\rm{recycle}}(n)$ and $f_{\rm{recycle}}=1$.}
\label{fig:corrplot}
\end{center}
\end{figure}
\begin{figure} 
\begin{center}
\resizebox{8cm}{!}{\includegraphics{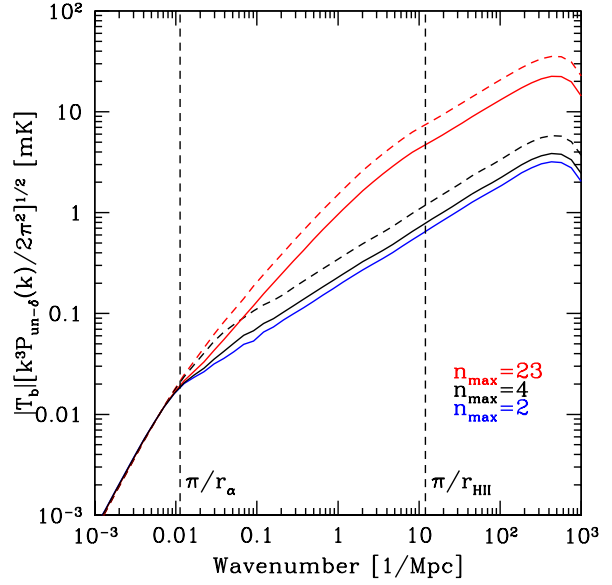}}\\%
\caption{Power spectrum for the Poisson fluctuations.  Line conventions are the same as in Figure \ref{fig:densityplot}.  Vertical lines indicate the scales corresponding to $2r_\alpha$ and $2r_{\rm{HII}}$.}
\label{fig:poissonpower}
\end{center}
\end{figure}

The top panel of Figure \ref{fig:corrplot} shows how the correlation function increases toward small scales.  This is a result of the $1/r^2$ dependence of the flux, which weights the correlations to small scales.  Including the \lyn resonances amplifies this, because the horizon scales skew the flux profile to small radii (see Figure \ref{fig:fluxprofile}).  On large scales the correlation function decreases as the two points A and B share fewer sources. For $r>2r_\alpha$ a single source cannot affect both points so $\xi_P=0$.  Including $f_{\rm{recycle}}(n)$ reduces $\xi_P$, especially on the smallest scales, because it decreases the efficiency of coupling from level $n$.  The bottom panel of Figure \ref{fig:corrplot} shows that the suppression on small scales is $(0.63)^2\approx0.40$, because $\xi_P$ depends on two powers of the flux.  Note that, on large scales, $\xi_P$ increases with the proper $f_{\rm{recycle}}$.  This results from the way we have normalised to $x_\alpha=1$, which reduces the flux of a given source on large scales as $f_{\rm{recycle}}$ increases (see Figure \ref{fig:fluxprofile}).  As with $W(k)$, the scale dependence is weak except on large scales, where rapid changes occur at the appropriate horizons.  Proper treatment of the recycling fractions is clearly necessary to understand the shape of $\xi_P$.

These features have similar effects on the power spectrum of fluctuations uncorrelated with the density fluctuations
\begin{equation}
P_{un-\delta}(k)\equiv P_{\mu^0}-\frac{P_{\mu^2}^2}{4P_{\mu^4}}=\left(\frac{x_\alpha}{\tilde{x}_{tot}}\right)^2P_{P}(k),
\end{equation}
as shown in Figure \ref{fig:poissonpower}.  On large scales, taking $f_{\rm{recycle}}=1$ slightly amplifies the power.  On small scales, they significantly reduce the power by $\approx60\%$.  They also affect the shape of the power spectrum, especially near $k=\pi/r_\alpha$, where the lack of $\rm{Ly}\beta\rightarrow\rm{Ly}\alpha$ imprints a knee on the power spectrum.  We also note that the sharp \lyn horizons imprint weak oscillations on the power spectrum, especially if $f_{\rm{recycle}}=1$ (though these will likely be smoothed by photon diffusion).

\subsection{Nonlinearities in the Wouthuysen-Field coupling}
\label{ssec:nonlinear}

To this point, we have used equation \eqref{deltatb} to compute the brightness
temperature fluctuations.  This assumes that all the underlying
perturbations are linear; obviously, at $x_\alpha=1$, this may only be
marginally satisfied.  When the radiation background is large, the
brightness temperature becomes insensitive to the coupling strength
and $P_{T_b}$ will be smaller than our estimate.  When will such
corrections become important?  One obvious test is whether the typical
fluctuation $T_b$ is comparable to the maximum brightness temperature
decrement between coupled and uncoupled gas, $\delta T$ (i.e., if
$T_s=T_k$ in eq. 1).  But nonlinearities may be important even if this condition is not satisfied. A universe with discrete strongly coupled regions
separated by uncoupled IGM could have small rms variations, even
though nonlinearities are extremely important in fixing the brightness
temperature of the strongly-coupled regions.

Instead we must look deeper at the nature of the fluctuations.  First,
note from Figure~\ref{fig:fluxprofile} that individual galaxies most
likely provide only weak coupling: $x_\alpha \ll 1$ except near to the
sources, at least if small galaxies (near the atomic cooling
threshold) are responsible for most of the radiation background.  This
is not surprising:  because of Olber's paradox, each logarithmic
radius interval contributes equally to the background flux in a 
homogeneous
universe.  The higher Lyman-series photons, together with the finite
speed of light, also do not dramatically increase the weighting on
nearby radii.  Thus we expect a substantial fraction of the flux to
come from large distances, where density fluctuations are weak.  This
immediately suggests that the density-dependent power spectrum
described in Section \ref{ssec:density} will not require substantial nonlinear
corrections.  

More quantitatively, the fluctuations become nonlinear when
$\delta_{x_\alpha}=W(k) \delta(k) \gtrsim 1$; thus we require
\begin{equation}
W(k) \gtrsim 1/[\sigma(R)\,D(z)],
\label{eq:nl1}
\end{equation}
where $\sigma(R)$ is the typical density fluctuation on scale $R \sim
1/k$.  Figure~\ref{fig:densityplot}\emph{b} shows, however, that $W(k)$ is of order
unity only for $k \lesssim 0.1$ Mpc$^{-1}$, where the density fluctuations
are themselves tiny at these redshifts.  Thus, we conclude that a
linear treatment is adequate for computing the $P_{\mu^2}$ power
spectrum, because it is primarily driven by large scale fluctuations.

The Poisson fluctuations in Section \ref{ssec:poisson} are more problematic.  By
definition, 
\begin{equation}
\xi_P({ r_A}-{ r_B}) = \langle x_\alpha({ r_A}) \,
x_\alpha({ r_B}) \rangle/x_\alpha^2 - 1.
\label{eq:nl2}
\end{equation}
Thus, on scales at which $\xi_P \gtrsim 1$, the radiation background near
galaxy overdensities on this scale is considerably larger than its
average value, indicating that nonlinear effects are important.  In
the particular model we have examined, $\xi_P$ is large only on
comoving scales $\lesssim 10$ kpc, so nonlinear effects are again
negligible.  However, the amplitude of the Poisson fluctuations
increases rapidly as the source density decreases: in models with
fewer sources at $x_\alpha=1$, or which strongly weight massive
galaxies, nonlinearities may be important.  A maximum value to the
variance on any scale is $\delta T^2 Q (1-Q) < 0.25\, \delta T^2$, where
$Q$ is the volume filling factor of regions with $x_\alpha \gtrsim 1$.  In
particular, a linear treatment for sources with $f_\star \propto m^{2/3}$ can violate this
limit on scales $k \gtrsim 1$ Mpc (see, e.g., Figs.~5 and 7 of BL05); in those cases the observed fluctuations can be much weaker than linear theory predicts (though it will also be non-gaussian).
\section{Conclusions}
\label{sec:conclusions}
In this paper, we have explored the effects on the spin-kinetic temperature coupling of photons that redshift into Lyman resonances.  First, we considered the effect of direct coupling via resonant scattering of \lyn photons.  We showed that the possibility of cascades greatly reduces the number of times a \lyn photon scatters before escaping.  Consequently, the coupling is negligible and may be correctly ignored.  A side effect of the reduced scattering rate is to make the \lyn contribution to IGM heating extremely small, even if $T_n$ is not in equilibrium with $T_k$.

Next we considered the increased \lya flux that results from atomic cascades.  Following the selection rules and transition rates, we calculated the probability that a \lyn photon is converted into \lya and showed that $f_{\rm{recycle}}\rightarrow0.36$ as $n$ increases.  This is significantly smaller than the value $f_{\rm{recycle}}=1$ usually assumed [e.g., by BL05].  For a typical Pop. III source spectrum, we showed that only $63\%$ of the emitted photons will be converted into \lyans, delaying the onset of coupling ($x_\alpha=1$) for a given set of source parameters.  

Incorporating the correct $f_{\rm{recycle}}$ modifies the flux profile of an individual source and reduces the coupling on small scales by about a factor of $3$ for fixed source parameters.  In addition, the cascade process imprints discontinuities onto the flux profile.  Using the correct atomic physics reduces the amplitude of these discontinuities and removes one due to the \lyb resonance.  Unfortunately, their weakness is likely to frustrate attempts to use these discontinuities as a standard ruler.

We then recalculated the power spectra of BL05, incorporating the correct $f_{\rm{recycle}}$.  This showed a reduction in power of $\sim37\%$ (Figure \ref{fig:densityplot}) on intermediate scales for density correlated fluctuations and of $\sim 64\%$ (Figure \ref{fig:poissonpower}) on small scales for fluctuations uncorrelated with the density.  It is possible to mimic this loss of power by changing the shape of the stellar spectrum.  On small scales, a reduction in the star formation rate will produce a similar reduction in flux.  Incorporating the proper $f_{\rm{recycle}}(n)$ is thus crucial to correctly interpreting 21 cm observations near the time of first light.  On the other hand, the effects that we have described become unimportant once \lya coupling saturates so that $T_s\rightarrow T_k$.  In this regime, fluctuations in the \lya flux have little effect on the power spectrum and perturbations in the density and reionization fraction dominate.

Several experiments, including LOFAR, MWA, PAST, and the future SKA\footnote{See http://www.skatelescope.org/.}, are aiming to detect 21cm fluctuations of the type we have discussed here.  Hopefully, they will be able to study the first sources of light through their effect on the IGM around them.  The details of \lya coupling will determine the observability of this epoch.

We thank M.~Kamionkowski for helpful discussions and also our anonymous referee for several helpful comments during the revision process.  This work was supported in part by DoE DE-FG03-920-ER40701.


\appendix
\section{Calculating Einstein $A$ coefficients in the hydrogen atom}
\label{app:einsteinA}
The simplicity of the hydrogen atom permits us to compute matrix elements for radiative transitions analytically.  We are interested in the Einstein $A$ coefficients, which may be written as  \citep{sobelman1972}
\begin{multline}
A(n,l,j,n',l',j') \\= \frac{64\pi^2}{3 h\lambda^3}e^2(2j'+1)\left\{ \begin{array}{ccc}
l & j & 1/2 \\
j' & l' & 1 
\end{array}
\right\}^2 l_> (R^{n',l'}_{n,l})^2.
\end{multline}
Here $\{\ldots\}$ denotes the Wigner $6-j$ symbol and we assume spin-1/2 particles.
In this expression, the matrix element $R^{n',l'}_{n,l}$ has the usual quantum numbers $n$, $l$, and $n_r$ with $n-l-1\equiv n_r$.  We use the sets $(n,n_r,l)$ and $(n',n_r',l')$ to describe the upper and lower levels in the spontaneous transition.  In addition, we let $l_>$ be the greater of $l$ and $l'$. Because they are separated by $\Delta l=l'-l=\pm1$, $2l_>=l+l'+1$.  

We next define, for $n'-l_>\ge1$,
\begin{equation}
r=n'-l_>-1\ge 0.
\end{equation}
We also let
\begin{equation}
u=(n-n')/(n+n');
\end{equation}
\begin{equation}
v=1-1/u^2=-4n n'/(n-n')^2;
\end{equation}
\begin{equation}
w=v/(v-1)=4nn'/(n+n')^2.
\end{equation}
Thus we only need $R^{n',l'}_{n,l}\equiv a_0 R$, where $a_0$ is the Bohr radius.  This is given by \citep{rudnick1935}
\begin{multline}
R=2^{l+l'+4}n^{l'+3}n'^{l+3}\frac{(n-n')^{n-n'-1}}{(n+n')^{n+n'+1}}\\ \times  \left[\frac{(n+l)!}{(n'+l')!(n-l-1)!(n'-l'-1)!}\right]^{1/2}P_{\pm}.
\end{multline}
Here $P_-$ and $P_+$ (the subscripts corresponding to the sign of $\Delta l$) are  terminating hypergeometric series
\begin{multline}
2n' P_-=(-1)^r\left[(n-n')^{2r}(2l_>+r)!/(2l_>)!\right] \\ \times\big[(n+n')F(-r,-n+l_>+1;2l_>+1;v)\\-(n-n')F(-r,-n+l_>;2l_>+1;v)\big],
\end{multline}
\begin{multline}
2n P_+=(-1)^r\left[(n-n')^{2r}(2l_>+r)!/(2l_>)!\right] \\ \times \big[(n+n')(n-l_>)F(-r,-n+l_>+1;2l_>+1;v)\\-(n-n')(n+l_>)F(-r,-n+l_>;2l_>+1;v)\big].
\end{multline}
These expressions can be inserted into equations \eqref{eqn:atomprob} and \eqref{f_recycle_iterate} to compute $P_{if}$ and $f_{\rm{recycle}}(n)$. They are in good agreement with existing experimental measurements\footnote{See reference data at the National Institute of Standards and Technology, http://physics.nist.gov/.}. 


\begin{thebibliography}{}

\bibitem[\protect\citeauthoryear{{Allison} \& {Dalgarno}}{{Allison} \&
  {Dalgarno}}{1969}]{allison1969}
{Allison} A.~C.,  {Dalgarno} A.,  1969, \apj, 158, 423

\bibitem[\protect\citeauthoryear{{Barkana} \& {Loeb}}{{Barkana} \&
  {Loeb}}{2005a}]{bl2005sep}
{Barkana} R.,  {Loeb} A.,  2005a, \apjl, 624, L65

\bibitem[\protect\citeauthoryear{{Barkana} \& {Loeb}}{{Barkana} \&
  {Loeb}}{2005b}]{bl2005detectgal}
{Barkana} R.,  {Loeb} A.,  2005b, \apj, 626, 1 [BL05]

\bibitem[\protect\citeauthoryear{{Bharadwaj} \& {Ali}}{{Bharadwaj} \&
  {Ali}}{2004}]{bharadwaj2004}
{Bharadwaj} S.,  {Ali} S.~S.,  2004, \mnras, 352, 142

\bibitem[\protect\citeauthoryear{{Breit} \& {Teller}}{{Breit} \&
  {Teller}}{1940}]{breit1940}
{Breit} G.,  {Teller} E.,  1940, \apj, 91, 215

\bibitem[\protect\citeauthoryear{{Chen} \& {Miralda-Escud{\' e}}}{{Chen} \&
  {Miralda-Escud{\' e}}}{2004}]{chen2004}
{Chen} X.,  {Miralda-Escud{\' e}} J.,  2004, \apj, 602, 1

\bibitem[\protect\citeauthoryear{{Ciardi} \& {Madau}}{{Ciardi} \&
  {Madau}}{2003}]{ciardi2003}
{Ciardi} B.,  {Madau} P.,  2003, \apj, 596, 1

\bibitem[\protect\citeauthoryear{{Field}}{{Field}}{1958}]{field1958}
{Field} G.~B.,  1958, Proc.~I.~R.~E., 46, 240

\bibitem[\protect\citeauthoryear{{Field}}{{Field}}{1959a}]{field1959spint}
{Field} G.~B.,  1959a, \apj, 129, 536

\bibitem[\protect\citeauthoryear{{Field}}{{Field}}{1959b}]{field1959relax}
{Field} G.~B.,  1959b, \apj, 129, 551

\bibitem[\protect\citeauthoryear{{Furlanetto}, {Schaye}, {Springel} \&
  {Hernquist}}{{Furlanetto} et~al.}{2005}]{furlane2005structure}
{Furlanetto} S.~R.,  {Schaye} J.,  {Springel} V.,    {Hernquist} L.,  2005,
  \apj, 622, 7

\bibitem[\protect\citeauthoryear{{Furlanetto}, {Sokasian} \&
  {Hernquist}}{{Furlanetto} et~al.}{2004}]{fsh2004}
{Furlanetto} S.~R.,  {Sokasian} A.,    {Hernquist} L.,  2004, \mnras, 347, 187

\bibitem[\protect\citeauthoryear{{Furlanetto}, {Zaldarriaga} \&
  {Hernquist}}{{Furlanetto} et~al.}{2004}]{fzh2004}
{Furlanetto} S.~R.,  {Zaldarriaga} M.,    {Hernquist} L.,  2004, \apj, 613, 1

\bibitem[\protect\citeauthoryear{{Gunn} \& {Peterson}}{{Gunn} \&
  {Peterson}}{1965}]{gp1965}
{Gunn} J.~E.,  {Peterson} B.~A.,  1965, \apj, 142, 1633

\bibitem[\protect\citeauthoryear{{Hirata}}{{Hirata}}{2005}]{hirata2005}
{Hirata} C.~M.,  2005, \mnras, submitted (astro-ph/0507102)

\bibitem[\protect\citeauthoryear{{Hogan} \& {Rees}}{{Hogan} \&
  {Rees}}{1979}]{hogan_rees1979}
{Hogan} C.~J.,  {Rees} M.~J.,  1979, \mnras, 188, 791

\bibitem[\protect\citeauthoryear{{Loeb} \& {Rybicki}}{{Loeb} \&
  {Rybicki}}{1999}]{loeb_rybicki1999}
{Loeb} A.,  {Rybicki} G.~B.,  1999, \apj, 524, 527

\bibitem[\protect\citeauthoryear{{Loeb} \& {Zaldarriaga}}{{Loeb} \&
  {Zaldarriaga}}{2004}]{loeb_zald2004_21cm}
{Loeb} A.,  {Zaldarriaga} M.,  2004, Physical Review Letters, 92, 211301

\bibitem[\protect\citeauthoryear{{Madau}, {Meiksin} \& {Rees}}{{Madau}
  et~al.}{1997}]{mmr1997}
{Madau} P.,  {Meiksin} A.,    {Rees} M.~J.,  1997, \apj, 475, 429

\bibitem[\protect\citeauthoryear{{Morales} \& {Hewitt}}{{Morales} \&
  {Hewitt}}{2004}]{morales_hewitt2004}
{Morales} M.~F.,  {Hewitt} J.,  2004, \apj, 615, 7

\bibitem[\protect\citeauthoryear{{Naoz} \& {Barkana}}{{Naoz} \& {Barkana}}{2005}]{naoz2005}
{Noaz} S., {Barkana} R., 2005, \mnras, 362, 1047

\bibitem[\protect\citeauthoryear{{Pen}, {Wu} \& {Peterson}}{{Pen}
  et~al.}{2005}]{pen2005past}
{Pen} U.~L.,  {Wu} X.~P.,    {Peterson} J.,  2005, ChJAA, submitted
  (astro-ph/0404083)

\bibitem[\protect\citeauthoryear{{Press} \& {Schechter}}{{Press} \&
  {Schechter}}{1974}]{ps1974mfn}
{Press} W.~H.,  {Schechter} P.,  1974, \apj, 187, 425

\bibitem[\protect\citeauthoryear{{Rudnick}}{{Rudnick}}{1935}]{rudnick1935}
{Rudnick} P.,  1935, Phys. Rev., 114, 114

\bibitem[\protect\citeauthoryear{{Santos}, {Cooray} \& {Knox}}{{Santos}
  et~al.}{2005}]{santos2005}
{Santos} M.~G.,  {Cooray} A.,    {Knox} L.,  2005, \apj, 625, 575

\bibitem[\protect\citeauthoryear{{Scott} \& {Rees}}{{Scott} \&
  {Rees}}{1990}]{scott_rees1990}
{Scott} D.,  {Rees} M.~J.,  1990, \mnras, 247, 510

\bibitem[\protect\citeauthoryear{{Sheth} \& {Tormen}}{{Sheth} \&
  {Tormen}}{1999}]{sheth1999mfn}
{Sheth} R.~K.,  {Tormen} G.,  1999, \mnras, 308, 119

\bibitem[\protect\citeauthoryear{{Sobelman}}{{Sobelman}}{1972}]{sobelman1972}
{Sobelman} I.~I.,  1972, {Introduction to the Theory of Atomic Spectra}.
Pergamon Press, Oxford, UK

\bibitem[\protect\citeauthoryear{{Spergel} et~al.,}{{Spergel}
  et~al.}{2003}]{spergel2003wmap}
{Spergel} D.~N.,  et~al., 2003, \apjs, 148, 175

\bibitem[\protect\citeauthoryear{{Wouthuysen}}{{Wouthuysen}}{1952}]{wouth1952}
{Wouthuysen} S.~A.,  1952, \aj, 57, 31

\bibitem[\protect\citeauthoryear{{Zaldarriaga}, {Furlanetto} \&
  {Hernquist}}{{Zaldarriaga} et~al.}{2004}]{zfh2004freq}
{Zaldarriaga} M.,  {Furlanetto} S.~R.,    {Hernquist} L.,  2004, \apj, 608, 622

\bibitem[\protect\citeauthoryear{{Zheng}, {Kriss}, {Telfer}, {Grimes} \&
  {Davidsen}}{{Zheng} et~al.}{1997}]{zheng1997quasar}
{Zheng} W.,  {Kriss} G.~A.,  {Telfer} R.~C.,  {Grimes} J.~P.,    {Davidsen}
  A.~F.,  1997, \apj, 475, 469

\bibitem[\protect\citeauthoryear{{Zygelman}}{{Zygelman}}{2005}]{zygelman2005}
{Zygelman} B.,  2005, \apj, 622, 1356

\end{thebibliography}
 \end{document}